\documentclass[conference]{IEEEtran}
\IEEEoverridecommandlockouts
\usepackage{cite}
\usepackage{amsmath,amssymb,amsfonts}
\usepackage{algorithmic}
\usepackage{graphicx}
\usepackage{textcomp}
\usepackage{multirow}
\usepackage{caption}
\usepackage{tabularx}
\usepackage{subcaption}
\usepackage{xcolor}
\usepackage{algorithm}
\usepackage{booktabs}

\usepackage{array}       

\def\BibTeX{{\rm B\kern-.05em{\sc i\kern-.025em b}\kern-.08em
    T\kern-.1667em\lower.7ex\hbox{E}\kern-.125emX}}
\begin{document}

\title{Grid Frequency Stability Support Potential of Data Center: A Quantitative Assessment of Flexibility\\}

\author{
    \IEEEauthorblockN{
        Pengyu Ren\IEEEauthorrefmark{1}, 
        Wei Sun\IEEEauthorrefmark{1}, 
        Yifan Wang\IEEEauthorrefmark{1}, 
        Gareth Harrison\IEEEauthorrefmark{1}
    }
    \IEEEauthorblockA{
        \IEEEauthorrefmark{1}School of Engineering, University of Edinburgh, Edinburgh, Scotland\\
    }
    \IEEEauthorblockA{
        Email: s2192464@ed.ac.uk, w.sun@ed.ac.uk, s2154060@ed.ac.uk, gareth.harrison@ed.ac.uk
    }
}

\maketitle

\begin{abstract}
The rapid expansion of data center infrastructure is reshaping power system dynamics by significantly increasing electricity demand while also offering potential for fast and controllable flexibility. To ensure reliable operation under such conditions, the frequency-secured unit commitment (Safe UC) problem must be solved with enhanced modeling of demand-side frequency response. In this work, we propose a data-driven linearization framework based on decision tree–based constraint learning (DT-CL) to embed nonlinear nadir frequency constraints into mixed-integer linear programming (MILP). This approach enables tractable co-optimization of generation schedules and fast frequency response (FFR) from data centers. Through case studies on both a benchmark system and a 2030 future scenario with higher DC penetration, we demonstrate that increasing the proportion of flexible DC load consistently improves system cost efficiency and supports renewable integration. However, this benefit exhibits diminishing marginal returns, motivating the introduction of the Marginal Flexibility Value (MFV) metric to quantify the economic value of additional flexibility. The results highlight that as DCs become a larger share of system load, their active participation in frequency response will be increasingly indispensable for maintaining both economic and secure system operations.
\end{abstract}


\section{Introduction}

The rise of artificial intelligence (AI), big data, and cloud computing has significantly driven the growth of data centers in recent years. According to the International Energy Agency (IEA), global data center electricity consumption reached 415 TWh in 2024—already 132\% of the UK’s total electricity use that year—and is projected to more than double to 945 TWh by 2030 \cite{IEA2025}. This explosive growth has raised global concerns over the sustainability and grid compatibility of data center operations, especially under net-zero carbon targets. Unlike conventional residential loads, the rapid pace and geographic concentration of data center development poses unique challenges for local electricity infrastructure \cite{epri2024datacenter}.

\begin{figure}[h]
\centering
\includegraphics[width=0.5\textwidth]{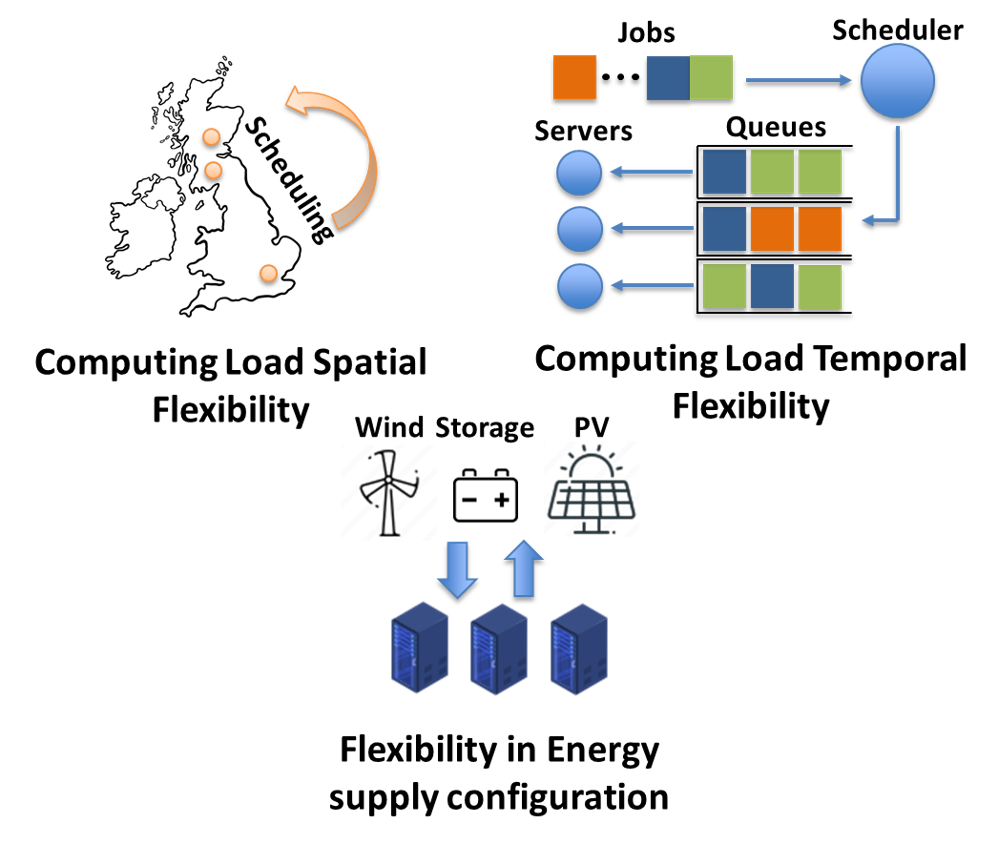}
\caption{Strategies for enabling data center flexibility}
\label{fig:Flex}
\end{figure}

Amid this trend, increasing attention has been given to the flexibility potential of data centers—the ability to modulate workload and electricity demand in response to grid signals. \cite{zhou2024ai} shows that AI-focused, GPU-heavy high-performance computing (HPC) data centers can provide power system flexibility at significantly lower cost than traditional CPU-based HPC centers, highlighting their potential as cost-effective resources for grid balancing. Figure~\ref{fig:Flex} illustrates recent developments in data center flexibility. Current strategies include dynamically shifting workloads in time and location, coordinating resource allocation across multiple cloud service providers, and aligning computing tasks with renewable energy availability.

In recent years, a growing body of literature has explored both the temporal and spatial flexibility of data centers, recognising their potential to provide grid support services such as load shifting\cite{li2014modeling} and fast frequency response\cite{zhang2021hpc}. For example, \cite{ding2025data} investigates how data centers can reduce their total electrical consumption costs through workload scheduling while maintaining Quality of Service (QoS) guarantees. Similarly, \cite{loganathan2017energy} proposes an energy-efficient job scheduling algorithm in cloud data centers by classifying jobs and applying preemption policies to maximize active host utilization and minimize the number of active physical machines, achieving up to 46\% energy savings compared to non-energy-aware baselines. \cite{sun2024privacy} proposes a privacy-preserving federated reinforcement learning framework for collaborative job scheduling across cloud providers, enabling energy sharing through decentralized decision-making while protecting operational privacy. \cite{yang2021carbon} proposes a spatio-temporal workload migration mechanism that reduces carbon emissions by flexibly shifting tasks across geographically distributed data centers to match multi-regional renewable energy availability. 


However, most of these studies adopt the perspective of data center operators, focusing on minimizing electricity bills or improving internal energy efficiency through workload scheduling \cite{ding2025data, loganathan2017energy, sun2024privacy}. A few recent works have begun to explore the system-level benefits of spatial and temporal workload shifting, such as improving renewable energy utilization \cite{yang2021carbon, zheng2020mitigating}, reducing grid violations \cite{chen2021operational}, and lowering carbon emissions \cite{seyyedi2024application}. Nevertheless, these efforts typically rely on simulation-based analyses and do not integrate such flexibility into formal grid operation models. In particular, there remains a lack of security-aware dispatch models that assess the impact of data center flexibility on frequency response and unit commitment decisions in real power systems.

To better understand the system-wide value of data center flexibility, it is crucial to shift the analytical perspective from isolated, facility-level optimizations to a holistic view of integrated power system operations \cite{chu2020towards}. A central challenge in this context lies in coordinating flexible loads—such as data centers—within established power system scheduling frameworks, particularly the Unit Commitment (UC) problem. In UC, the system operator determines the optimal dispatch and on/off status of generation units to minimize total operating costs. When frequency constraints are considered, the model must also ensure sufficient system inertia, typically requiring the commitment of certain thermal generators \cite{meng2019fast}.

Several studies have extended the UC framework to incorporate frequency-related considerations. For instance, \cite{teng2015stochastic} proposes a stochastic UC model that integrates frequency constraints to assess the economic impact of frequency response capabilities. Similarly, \cite{li2021frequency} introduces a stochastic planning model that incorporates the frequency response behavior of wind power to ensure system frequency stability. In parallel, \cite{lahon2020impact} explores the role of large-scale data centers in the 2030 Irish power system, embedding their operations into a mixed-integer unit commitment formulation to study cost and flexibility trade-offs.

However, many existing models oversimplify frequency response dynamics, especially when dealing with multi-resourece frequency response like data center. Homogeneous response characteristics across all resources are often assumed, overlooking critical device-specific factors such as activation delays, ramping behavior, and capacity constraints. These limitations may lead to inaccurate assessments of flexible resources’ true value and hinder the development of effective frequency support strategies. To fill in the above research gap, this paper proposes an innovative safe UC method based on decision tree linearization. In detail, the main contributions of this paper are summarized as follows:

\begin{itemize}
\item Quantitatively evaluate the value of data center frequency flexibility in improving the reliability of the power system under high penetration of renewables energy. By enabling data centers to provide fast frequency response (FFR), the proposed framework reduces overall system operation costs and mitigates reliance on conventional thermal generators, thereby supporting a more efficient and sustainable power system operation.

\item Propose a safety-aware unit commitment (Safe UC) model that integrates data-driven frequency safety constraints derived from historical operation data. We introduce a novel constraint generation mechanism using a logistic-regression-based slope tree, which partitions the operational space into safe and unsafe regions. The resulting piecewise linear inequalities are directly embedded in the MILP formulation, enabling interpretable and computationally tractable enforcement of empirical safety boundaries.

\item Demonstrate the effectiveness of the proposed approach on a high-renewable test system. Results show that incorporating data center frequency response improves renewable utilization, reduces curtailment, and contributes to secure operation under increasing system uncertainty, offering practical insights for future grid planning and data center participation.
\end{itemize}


\section{Safe unit commitment modelling}

\begin{figure}[htbp]
\centerline{\includegraphics[width = 0.5\textwidth, height = 0.38\textwidth]{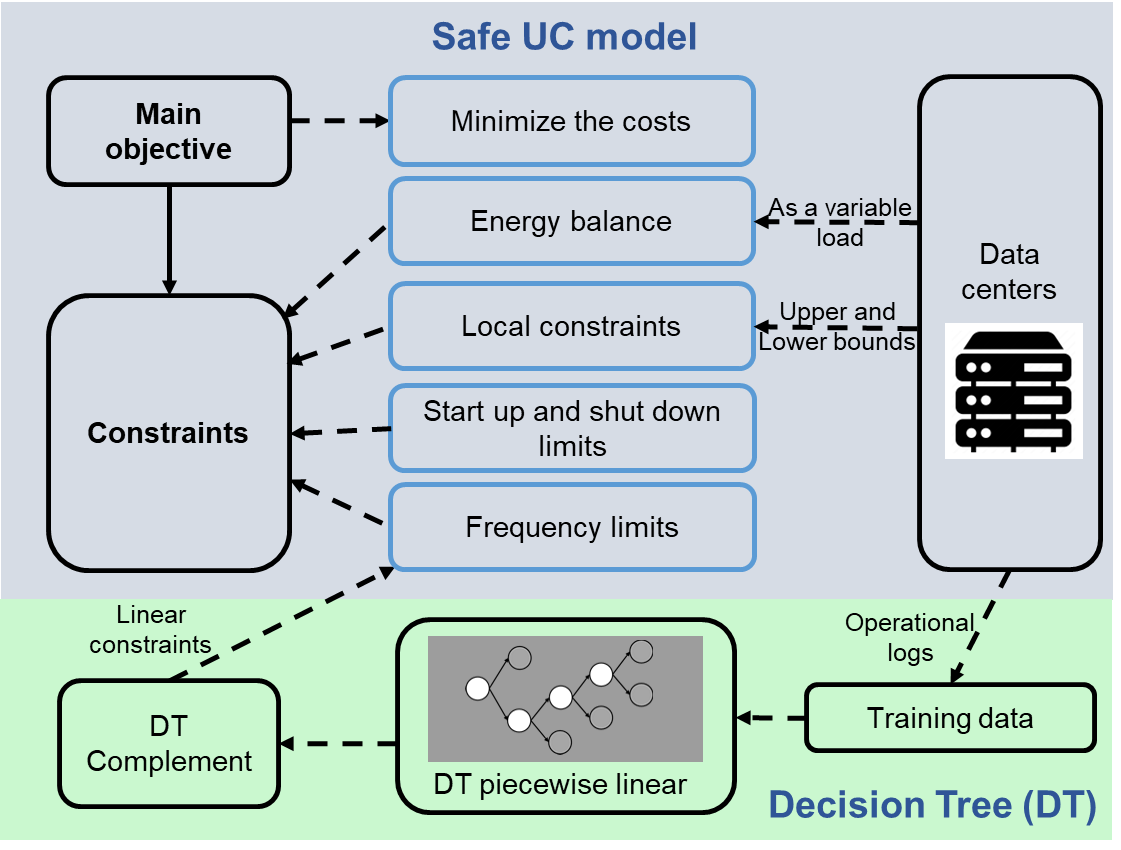}}
\caption{Outline of Safe UC modelling.}
\label{fig2}
\end{figure}

To address the challenges of integrating flexible data center demand into the UC problem, we propose a hybrid framework that combines traditional optimization modeling with data-driven decision-making. As illustrated in Fig.~\ref{fig2} our method augments a conventional UC formulation with a decision tree (DT)-based constraint learning mechanism to improve operational safety and adaptability under uncertainty.

The Safe UC model lies at the core of the framework. Its primary objective is to minimize total generation cost, subject to a range of physical and operational constraints including energy balance, frequency limits, local network limitations, and start-up/shut-down dynamics. Data centers are modeled as flexible and controllable loads, whose consumption is treated as a variable within specified upper and lower bounds. These bounds, derived from operational capabilities, are incorporated into the local constraints of the UC formulation.

To enhance safety and robustness, we introduce data-driven safety constraints learned from historical operational logs. Specifically, we train a decision tree classifier on past system data to identify safe vs. unsafe operating regions. The resulting tree is then converted into a piecewise linear form, which is further translated into a set of linear constraints compatible with mixed-integer linear programming (MILP) formulations. 

The DT complement module, as shown in the lower part of Fig.~\ref{fig2}, serves as a constraint-generation engine. It translates empirical knowledge into constraints that can be seamlessly combined with the analytical UC model. In doing so, the overall framework maintains interpretability while leveraging real-world data to improve decision quality, particularly under scenarios involving high demand variability and renewable generation uncertainty.

\subsection{Defining data center flexibility}

Before formulating the safe unit commitment model, it is essential to define the nature of flexibility that data centers can provide to the power system. As discussed in the introduction part, numerous studies have demonstrated that data centers possess inherent flexibility due to their internal structure and the nature of computing workloads they handle.

The degree of controllability largely depends on the types of incoming jobs, which can be broadly divided into:

\begin{itemize}
    \item \textbf{Delay-sensitive (real-time) workloads}, which must be executed immediately upon arrival to meet user expectations or quality-of-service requirements.
    
    \item \textbf{Delay-tolerant (batch) workloads}, which can be scheduled flexibly within a given time window, provided they complete before a specified deadline.
\end{itemize}

In general, batch workloads are the primary source of demand-side flexibility in data centers. These tasks can be modulated, deferred, or even interrupted without violating service-level agreements. This flexibility enables data centers to adapt their power consumption in response to system needs.

In this study, since our focus is on enabling FFR, only workloads that can be rapidly deferred—typically batch jobs with high schedulability—are considered. Other forms of flexibility, such as long-term workload peak shifting or modulation of cooling system loads, are beyond the scope of this work and not included in the model.

\subsection{Objective function}

The objective of the unit commitment model with fast frequency response is to minimise the total operating cost over a scheduling horizon:


\begin{equation}
\footnotesize
\min \sum_{t \in \mathcal{T}} \left[ 
\underbrace{
\sum_{i \in \mathcal{I}} \left( \beta_1 P_{i,t}^2 + \beta_2 P_{i,t} + \beta_3 \right)
}_{\text{Generator costs}}
+
\underbrace{
\gamma_1 R^{\mathrm{DC}}_t + \gamma_2 (R^{\mathrm{DC}}_t)^2
}_{\text{DC FFR costs}}
\right]
\end{equation}

where \( P_{i,t} \) is the active power output of generator \( i \) at time step \( t \), \( R^{\mathrm{DC}}_t \) is the DC fast frequency response potential at time \( t \),
\( \beta_1, \beta_2, \beta_3 \) are generation cost coefficients, \( \gamma_1, \gamma_2 \) are linear and quadratic cost coefficients for DC response.

The formulation balances two key components of system operation: the economic efficiency of power generation and the provision of frequency regulation services by flexible demand resources. The generator cost term reflects the traditional economic dispatch, where generation units are scheduled based on their cost curves. In contrast, the DC FFR cost term quantifies the economic burden of engaging data centers in frequency support, penalizing excessive reliance on demand flexibility through both fixed and escalating marginal costs.

Importantly, the inclusion of \( R^{\mathrm{DC}}_t \) as a decision variable bridges the physical UC optimization with the data-driven safety constraints. These variables are not only subject to operational bounds, but are also constrained by learned safe-operating regions derived from historical data via the DT mechanism. 

This coupling ensures that the optimisation respects both physical grid limits and empirically derived safety margins, enhancing the realism and robustness of the scheduling decisions. As such, the objective function lies at the core of a hybrid paradigm that marries analytical modeling with machine learning–enabled constraint generation, enabling secure and cost-effective operation under uncertainty from renewable generation and flexible loads.

\subsection{Conventional Unit commitment constraints}
In addition to the frequency considerations discussed later, the core unit commitment (UC) model is governed by a set of traditional operational constraints that ensure the physical feasibility and reliability of the system. These include power balance, generator capacity limits, ramping limitations, and minimum up/down time requirements, as described below:

\begin{align}
&\sum_{i} P_{i,t} + \sum_{i} P^{\text{wind}}_{it} = P^{\text{curt}}_t + P^{\text{DC}}_t + P^{\text{load}}_t \\
&P^{\min}_i u_{i,t} \leq P_{i,t} \leq P^{\max}_i u_{i,t} \\
&P_{i,t} - P_{i,t-1} \leq \rho^{up}_i u_{i,t} \\
&P_{i,t-1} - P_{i,t} \leq \rho^{up}_i u_{i,t} \\
&1 - u_{i,t} \geq \sum_{d=1}^{\delta_i} (1 - u_{i,t-d})
\end{align}
where \( P^{\text{wind}}_{it} \) represents wind power generation; \( P^{\text{curt}}_t \) is the curtailed power; \( P^{\text{DC}}_t \) is the power consumed by data center loads; \( P^{\text{load}}_t \) is the total system demand; \( P^{\min}_i \) and \( P^{\max}_i \) are the minimum and maximum generation limits of unit \( i \), respectively; \( \rho^{up}_i \) and \( \rho^{up}_i \) denote the generator’s ramp-up and ramp-down limits; and \( \delta_i \) is the minimum down-time duration required once unit \( i \) is turned off.

Each constraint serves a specific operational or stability requirement. Equation (2) ensures active power balance between generation and demand (including curtailed and DC load). Equation (3) limits generator outputs based on commitment status. Equation (4)-(5) enforce generator ramp-up and ramp-down rate limits. Equation (6) ensures that once a generator is turned off, it must remain off for a minimum down-time duration \( \delta_i \).

\subsection{Frequency response modelling}\label{AA}
While the above constraints define the traditional operational boundaries of the UC problem, they do not explicitly capture the system's dynamic behavior during frequency disturbances. In modern grids with high renewable penetration and demand-side flexibility, modelling frequency response characteristics becomes essential to ensure stability and system security \cite{jiang2024frequency}. In this context, we incorporate an explicit model of frequency response behavior from both conventional generators and data center loads, as described in the next section.

\begin{figure}[htbp]
\centerline{\includegraphics[width = 0.45\textwidth]{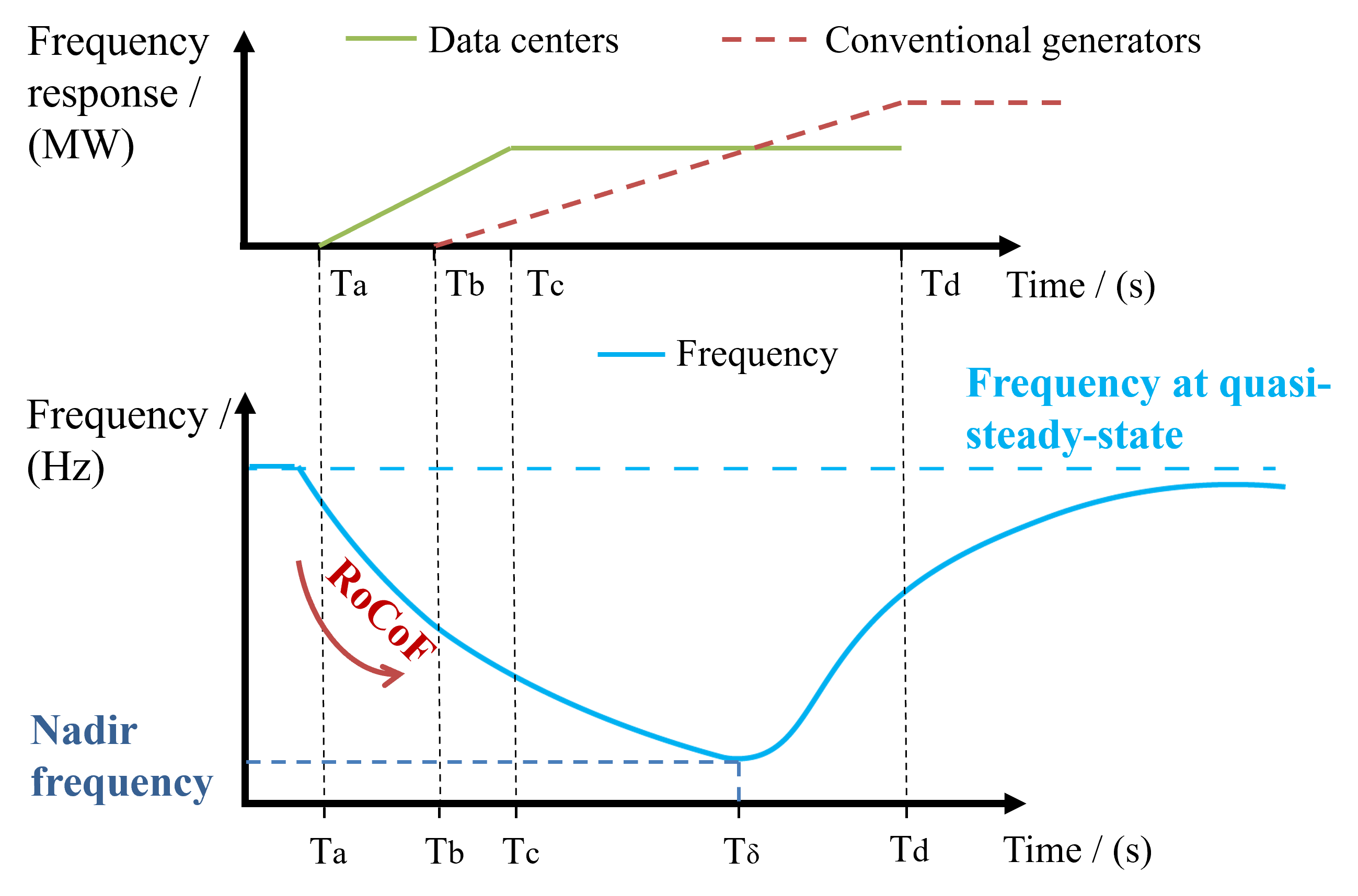}}
\caption{Assumed frequency response process of data center and conventional generators.}
\label{FRmodel}
\end{figure}

Suppose a power outage or unit dropout occurs, when the system starts to shift in frequency due to a power imbalance. When the frequency offset reaches the frequency deadband $\Delta f_{DB}$ set by the system, the ancillary service starts to response. As shown in Fig.~\ref{FRmodel}, The model assumes that the fast frequency response process of both data center and conventional generators is a ramp function, and they do not start responding at the same time.


The frequency constraint of the system can be delineated into three distinct components, as the Fig.~\ref{FRmodel} shows. The initial component imposes a constraint on the Rate of Change of Frequency (RoCoF) during a frequency variation, stipulating that the rate of change must not be excessively rapid to avert excessive stress on transmission lines. The second component encompasses a constraint on the frequency at nadir, ensuring that the system's frequency remains stable within an acceptable range. The final component comprises a constraint on the frequency at which the system attains a quasi-steady state following its frequency response. 

Commencing at time $T_{a}$, the frequency offset attains the deadband frequency threshold, prompting the DCs to initiate a ramping-type FFR. At time $T_{b}$, the conventional generator commences its FFR performance. At $T_{c}$, the FFR of the DC unit achieves its stable capacity and enters a maintenance phase. Subsequently, at $T_{d}$, the FFR of the conventional generator also reaches its stable capacity. This process is shown in Fig.~\ref{FRmodel}. The ramping functions for the two are shown below:

\begin{equation}
\Delta R_{DC}(t)=\left\{\begin{array}{ll}
0 & \text { if } t<T_{\mathrm{a}} \\
\frac{R_{DC}^{sta}}{T_{c}-T_{a}} \times \left(t-T_{\mathrm{a}}\right) & \text { if } T_{\mathrm{a}} \leq t \leq T_{\mathrm{c}} \\
R_{DC}^{sta} & \text { if } t \geq T_{\mathrm{c}}
\end{array}\right.
\end{equation}

\begin{equation}
\;\;\;\Delta R_{cg}(t)=\left\{\begin{array}{ll}
0 & \text { if } t<T_{\mathrm{b}} \\
\frac{R_{cg}^{sta}}{T_{d}-T_{b}} \times \left(t-T_{\mathrm{b}}\right) & \text { if } T_{\mathrm{b}} \leq t \leq T_{\mathrm{d}} \\
R_{cg}^{sta} & \text { if } t \geq T_{\mathrm{d}}
\end{array}\right.
\end{equation}

where the $\Delta R_{DC}(t)$ and the $\Delta R_{cg}(t)$ are FFRs of the DCs and conventional generators (basically are thermal units). The 
$R_{DC}^{sta}$ and the $R_{cg}^{sta}$ are maximum FFRs could be provided.
$T_{a}$ and $T_{b}$ are the start times of the FFRs of the DCs and conventional generators, respectively. Meanwhile, $T_{c}$ and $T_{d}$ are the times at which the DCs and conventional generators reach their maximum FFRs, respectively.

\subsection{Rate of change of frequency limit}\label{RoCoF}
In addition to modelling the ramping behavior of flexible resources, a critical aspect of frequency security lies in constraining the Rate of Change of Frequency (RoCoF) immediately following a contingency. RoCoF is a dynamic metric that reflects how rapidly the system frequency changes in response to power imbalances and is inversely proportional to the system’s total inertia. Excessive RoCoF may lead to protection system misoperations, equipment damage, or system instability.

The system inertia \( H_s \) can be expressed as:

\begin{equation}
H_{s}=\frac{\sum_{g \in \mathcal{G}} H_{g} \times P_{g}^{\max } \times B-\Delta P_{L}^{\max } \times H_{L}^{\max }}{f_{0}}
\end{equation}
where $H_{s}$ is the system inertia, $H_{g}$ is the inertia constant of each generator, $P_{g}^{\max }$ is the maximum output of each generator, $B$ is the binary control variable, $\Delta P_{L}^{\max }$ is the maximum loss of load, $H_{L}^{\max }$ is the inertia constant of load, and $f_{0}$ is the nominal frequency.

Accordingly, the system-wide RoCoF is formulated as:
\begin{equation}
\delta = \frac{\Delta P_{L}^{\max}}{2 H_{s}} \leq \delta_{\max}
\end{equation}
where $\delta$ denotes the instantaneous rate of frequency change and $\delta_{\max}$ is the permissible upper limit to ensure safe system operation.

\subsection{Nadir Frequency Constraint}\label{sec:nadir}

While the RoCoF constraint governs the initial rate of frequency deviation, frequency security must also account for the lowest frequency point reached after a disturbance—commonly referred to as the nadir. Following a loss of generation, the frequency continues to drop even after primary response is triggered, until sufficient energy is injected to restore balance. The nadir frequency thus depends on three key factors: system inertia, total primary frequency response, and the magnitude of the power imbalance.

To avoid triggering under-frequency load shedding (UFLS) or protection trips, a nadir constraint is imposed to ensure that system frequency stays above a critical threshold during transients. The time evolution of system frequency $\Delta f(t)$ can be described using the linearised swing equation \cite{huang2012sensitivity} :

\begin{equation}
2 H_s \frac{d \Delta f(t)}{dt} + D \cdot P_D \cdot \Delta f(t) = \sum_{g,s \in \mathcal{G}, \mathcal{S}} \Delta P_{g,s}(t) - \Delta P_{L}^{\max}
\end{equation}

where $D$ is the frequency damping coefficient, $P_D$ is the total demand, and $\Delta P_{g,s}(t)$ is the frequency response power from resource $g$ of type $s$.

By substituting the ramping expressions of $\Delta R_{\text{DC}}(t)$ and $\Delta R_{\text{cg}}(t)$ into this differential equation and solving for $\Delta f(t)$, we derive a four-phase nonlinear frequency trajectory. The nadir point, influenced by the interplay of inertia and response ramping delays, does not admit a closed-form linear constraint in standard MILP formulations. Therefore, we reformulate the nadir constraint using a data-driven linear approximation, which is detail discussed in Section. \ref{Nadirreform}. 


\subsection{Quasi-Steady-State (QSS) Constraint}

After the frequency nadir, the system enters a quasi-steady-state period where the frequency deviation is relatively constant. To maintain operational integrity and avoid secondary frequency control actions, the residual frequency deviation must not exceed a permissible value:

\begin{equation}
L_{\max} - R_t \leq \xi \cdot D^{\text{peak}} \cdot \lambda
\end{equation}

where $L_{\max}$ is the maximum disturbance size, $R_t$ is the total available frequency response at time $t$, $D^{\text{peak}}$ is peak demand, and $\xi$, $\lambda$ are empirical QSS coefficients derived from operational data.

\section{Reformulation of Nadir frequency limit}
\label{Nadirreform}
Considering that the frequency response characteristics of data centers differ from those of conventional generators, the original nadir frequency constraint becomes even more nonlinear. As a result, the nadir frequency limit cannot be directly incorporated into the MILP formulation of the MILP problem.  To overcome this, we apply a data-driven linearization technique based on decision trees.

\subsection{Decision tree linearization}

\begin{figure}[h]
    \centering
    \begin{subfigure}[t]{0.24\textwidth} 
        \centering
        \includegraphics[width=\textwidth]{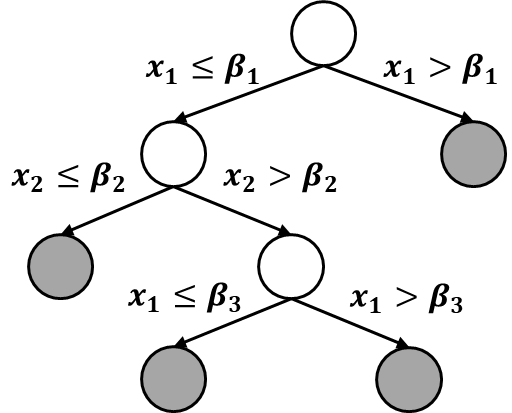}
        \caption{Classic decision tree topology.}
        \label{fig:dt}
    \end{subfigure}
    \hfill
    \begin{subfigure}[t]{0.24\textwidth} 
        \centering
        \includegraphics[width=\textwidth]{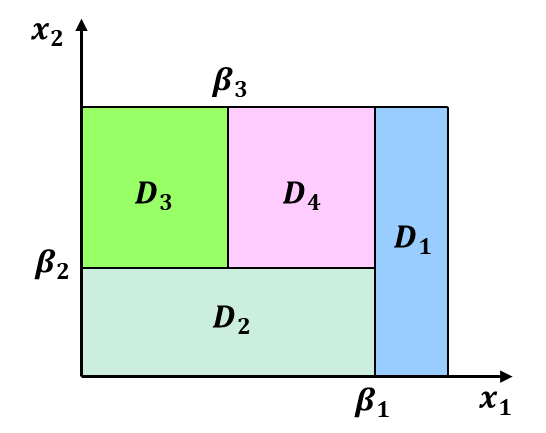}
        \caption{Domain partition.}
        \label{fig:dtrg}
    \end{subfigure}
    \caption{An illustration on a simple decision tree with splitting knots.}
    \label{fig:comparison}
\end{figure}

A decision tree is a supervised learning algorithm that recursively partitions the input space into subsets based on feature values, ultimately forming a tree-like structure where each leaf node represents a prediction, as shown in Fig. \ref{fig:dt}. When used for piecewise linearization of a nonlinear function, the decision tree algorithm divides the function's domain into multiple regions, as shown in Fig. \ref{fig:dtrg}, each of which can be approximated by a linear model.

DTs have been used in the power systes domain, including \cite{jia2024learning} who used a decision tree method to linearize the Q–V curve for voltage stability in power systems. In contrast, the present work conducts linearization in a three-dimensional space, due to the three-variable nature of the nonlinear frequency nadir constraints, which makes the process significantly more challenging.

\begin{figure}[h]
\centering
\includegraphics[width=0.5\textwidth]{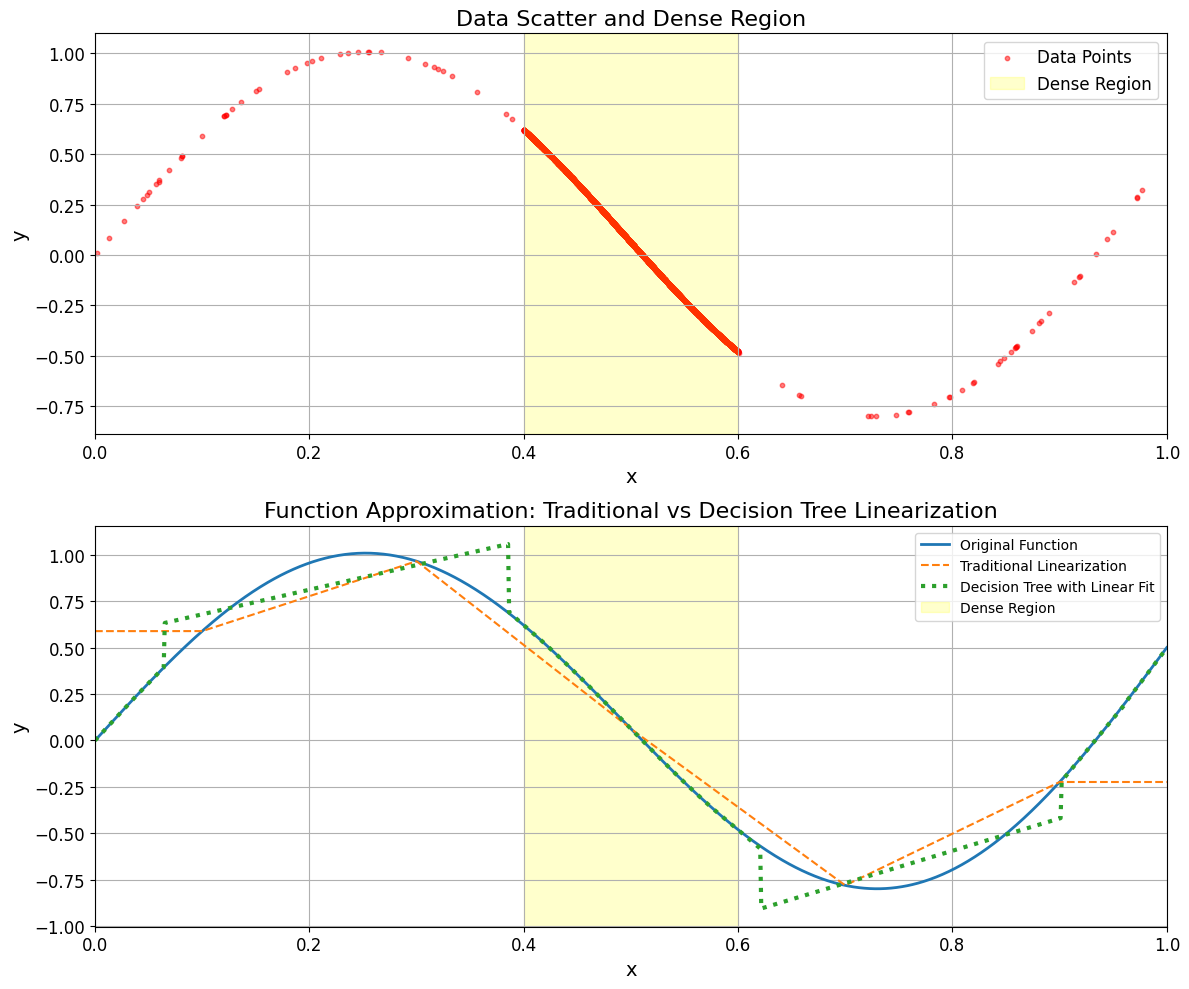}
\caption{Comparison between conventional piecewise linear and decision tree linear models.}
\label{fig:comparison}
\end{figure}


Figures \ref{fig:comparison} illustrate the differences between the linearization method using a Decision Tree and the traditional piecewise linearization approach. The upper plot shows a non-uniform sample of a sine curve, where the yellow region represents the densely sampled data points. As shown in the lower plot, the conventional piecewise linearization method adopts equally spaced segments to minimize the overall deviation between the final linearized expression and the original nonlinear relationship. In contrast, the Decision Tree captures the characteristics of the data. With the same number of segments, it sacrifices accuracy in less dense regions to ensure lower errors in areas with high-frequency data.

This approach provides an adaptive method for piecewise linearization, as the decision tree automatically determines the optimal partitions based on the data. However, the accuracy of the approximation depends on the tree's depth and splitting criteria, which control the granularity of the partitioning. Overly shallow trees may lead to coarse approximations, while overly deep trees risk overfitting and unnecessary complexity. The training of the decision tree, including parameter selection and interval adjustment, will be explained in detail in the next subsection.

\subsection{DT trainning algorithm}
To enable the training of the decision tree classifier, we first construct a Simulink-based frequency response simulation framework based on the analytical nadir frequency equation derived earlier. This control system mimics the dynamic behavior of frequency deviation following generation loss and includes both generator inertia and flexible load responses from data centers. For each unit commitment solution, we extract relevant features including the power output of conventional generators, the scheduled load level of data centers, and other system operating conditions. These are used as input variables to the Simulink model, while different data center response times are configured as system parameters.

The output of each simulation instance is a binary safety label, indicating whether the corresponding operating point satisfies the nadir frequency constraint under a specific disturbance. Collectively, this process generates a labeled dataset suitable for supervised learning, where each input vector represents an operational scenario and the label denotes its safety status.

To encode safety constraints directly from historical operation data, we construct a decision tree whose internal nodes are defined by logistic regression classifiers, referred to as a slope tree. Unlike conventional regression or classification trees that use axis-aligned splits, this structure allows each node to introduce a linear decision boundary, yielding a more compact and flexible partitioning of the input space.

Given a labeled dataset \((\mathbf{x}_i, y_i)\), where \(\mathbf{x}_i\) encodes the system's operating features and \(y_i \in \{0, 1\}\) indicates whether the operating point is safe or unsafe, the tree construction proceeds recursively as shown in Algorithm~\ref{algo1}. At each node, we fit a logistic regression model
\[
f(\mathbf{x}) = \mathbf{a}^\top \mathbf{x} + c
\]
to separate the two classes. The dataset is then split based on the sign of the score \(f(\mathbf{x})\): the left child receives points with \(f(\mathbf{x}) < 0\), and the right child receives those with \(f(\mathbf{x}) \geq 0\). The process stops when the maximum tree depth or minimum sample threshold is reached, or when all samples in the node belong to the same class.

\begin{algorithm}
\caption{BuildSlopeTree: Logistic Regression-Based Slope Decision Tree}
\label{algo1}
\begin{algorithmic}[1]
\REQUIRE Feature matrix $X$, label vector $y$, current depth $d$, max depth $d_{max}$, minimum samples $n_{min}$
\ENSURE Tree node or \textbf{None}

\IF{$d \geq d_{max}$ or $|X| < n_{min}$ or all $y_i$ are equal}
    \RETURN \textbf{None}
\ENDIF

\STATE Train logistic regression model $f(x) = \mathbf{a}^\top \mathbf{x} + c$ on $(X, y)$

\STATE Compute decision scores $\mathbf{s} \gets f(X)$

\STATE Partition data:
\begin{itemize}
  \item Left subset: $X_L, y_L \gets \{ (\mathbf{x}_i, y_i) \mid s_i < 0 \}$
  \item Right subset: $X_R, y_R \gets \{ (\mathbf{x}_i, y_i) \mid s_i \geq 0 \}$
\end{itemize}

\IF{$X_L$ or $X_R$ is empty}
    \RETURN \textbf{None}
\ENDIF

\STATE Recursively build left subtree: $T_L \gets \text{BuildSlopeTree}(X_L, y_L, d+1, d_{max}, n_{min})$
\STATE Recursively build right subtree: $T_R \gets \text{BuildSlopeTree}(X_R, y_R, d+1, d_{max}, n_{min})$

\IF{$T_L = T_R = \textbf{None}$}
    \RETURN \textbf{None}
\ENDIF

\RETURN Node $(\mathbf{a}, c, T_L, T_R)$
\end{algorithmic}
\end{algorithm}

\begin{algorithm}
\caption{CollectConditions: Traverse a slope tree to extract MILP-compatible linear inequalities}
\label{algo2}
\begin{algorithmic}[1]
\REQUIRE Tree root node $T$
\ENSURE List of linear inequalities $(\mathbf{a}, c)$

\IF{$T = \textbf{None}$}
    \RETURN empty list
\ENDIF

\STATE Initialize list $L \gets [(\mathbf{a}, c)]$
\STATE Append $CollectConditions(T.left)$ to $L$
\STATE Append $CollectConditions(T.right)$ to $L$
\RETURN $L$
\end{algorithmic}
\end{algorithm}

The DT complement module translates empirical safety knowledge into linear inequality constraints, which can be seamlessly embedded into MILP formulations. Unlike conventional regression or classification trees that use axis-aligned splits, the slope tree is a form of oblique decision tree, in which each internal node applies a logistic regression classifier to introduce a linear (oblique) decision boundary. This enables more compact and flexible partitioning of the input space, better aligning with the geometry of the nonlinear nadir frequency constraint.

The adaptivity of the tree structure enables data-driven refinement of safety boundaries. However, the depth and splitting criteria must be carefully chosen to balance approximation accuracy and computational tractability. Further details on the training procedure and constraint encoding will be provided in the following section.

However, the output of the decision tree is not immediately suitable for optimization solvers. To incorporate its results into the unit commitment problem, the learned decision rules must be reformulated into MILP-compatible linear inequalities. This is accomplished by traversing the tree using the procedure in Algorithm~\ref{algo2}, which collects the conjunction of node-level conditions along each path to a safe leaf. The resulting set of inequalities defines the feasible operating region in a form that can be directly embedded into the MILP formulation.

\subsection{Constraint embedding}

The resulting convex relationship between system frequency nadir and control variables can be expressed as a linear inequality. Specifically, this data-driven convex surface can be approximated using supervised learning techniques, allowing it to be formulated as part of the MILP model:

\begin{equation}
\theta_1 R^{\text{gen}}_t + \theta_2 R^{\text{DC}}_t + \theta_3 H_t + \theta_4 \geq 0
\end{equation}

Here, $\theta_1$, $\theta_2$, $\theta_3$, and $\theta_4$ are slope coefficients extracted via supervised learning using logistic regression–based decision trees. This transformation embeds the frequency nadir constraint as a linear inequality, enabling compatibility with MILP solvers used in unit commitment formulations. 

By doing so, the nonlinear frequency security boundary is effectively captured by a set of interpretable and computationally tractable constraints, integrating physics-informed machine learning into power system scheduling.

\section{Case study and analysis of results}
\begin{figure}[htbp]
\centerline{\includegraphics[width = 0.5\textwidth]{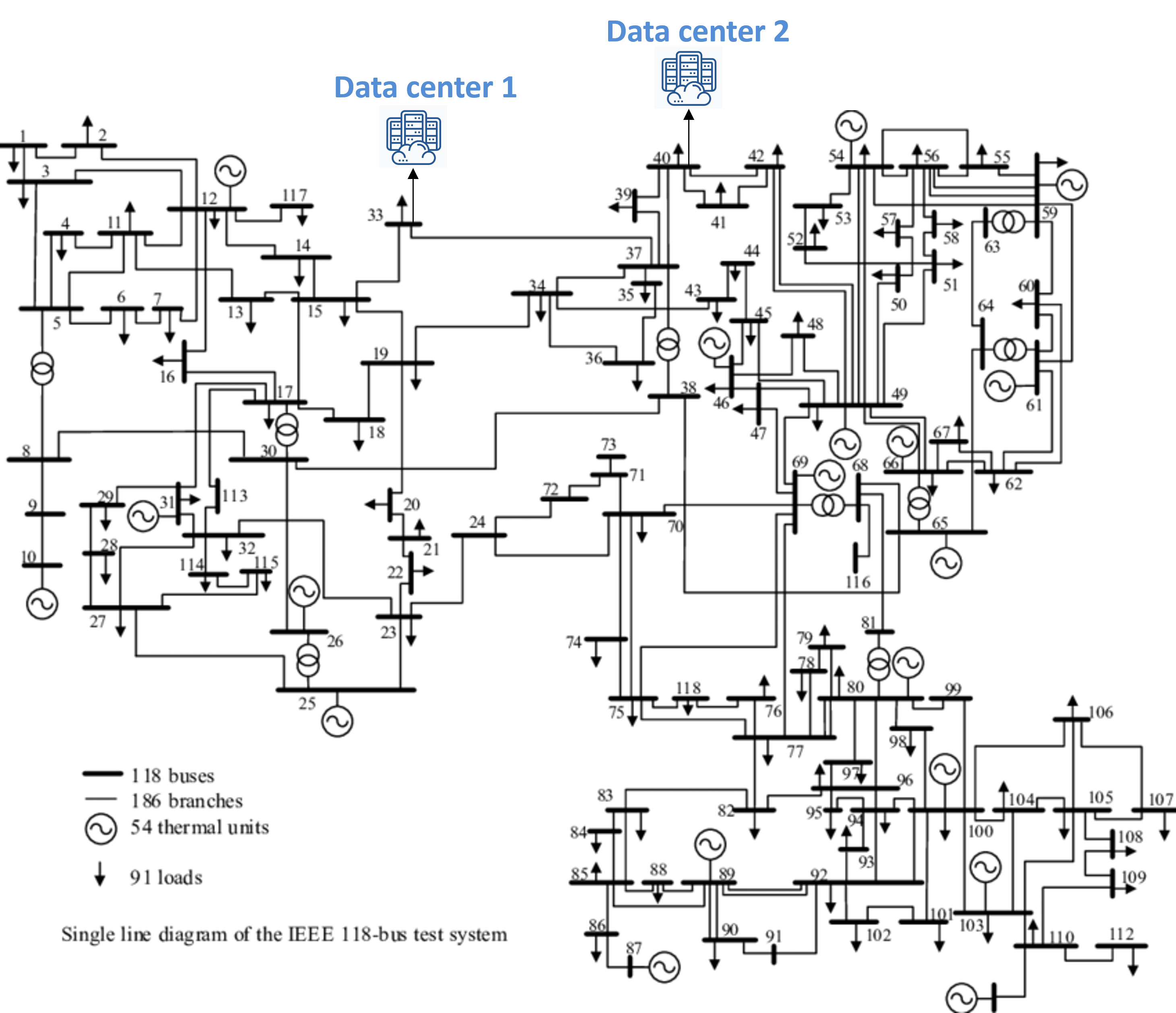}}
\caption{Modified IEEE 118 bus test system \cite{christie1993power}, including two data centers with 500MW peak demand each.}
\label{IEEE118}
\end{figure}

The power system dataset used in this case study is based on a modified IEEE 118-bus system \cite{christie1993power}, shown in Fig. \ref{IEEE118}, in which a high level of wind power penetration has been incorporated to simulate future power systems with renewable energy and data center integration. Several conventional generators have been replaced or augmented with wind farms located at strategically selected buses to reflect spatial diversity and variability. The optimization and solving language employed in this study is Julia 1.10.2. 

The system under study is configured with a total demand of 5242 MW, including 1000 MW from data center loads. Table~\ref{tab:sys_config} summarizes the installed capacities of conventional and renewable generation units, as well as the demand components. 

\begin{table}[h]
\centering
\caption{System Configuration Summary}
\label{tab:sys_config}
\begin{tabular}{lc}
\toprule
\textbf{Component} & \textbf{Capacity / Demand (MW)} \\
\midrule
Conventional Generators & 5437.69 \\
Wind Turbines           & 2718.84 \\
Data Center Load        & 1000.00 \\
Other System Load       & 4242.00 \\
\textbf{Total Load}     & \textbf{5242.00} \\
\bottomrule
\end{tabular}
\end{table}

The 1000 MW load of data center represents the peak rated capacity of the data center component. In the simulation model, this value serves as the upper bound of flexible load. Rather than assuming constant demand, the actual data center power consumption at each hour is derived from a year-long operational log covering 8760 hours from the year 2019. The normalized load profile (\( \alpha_t \)) is obtained from publicly available standard demand profiles developed by UK Power Networks, which provide hourly data for various demand types including data centers \cite{ukpn2024standardprofiles}. Specifically, the real-time load is modeled as a time-varying share of the peak, computed as:
\[
P_{\text{DC},t} = \alpha_t \cdot P_{\text{DC}}^{\max}, \quad \alpha_t \in [0, 1]
\]
where \( P_{\text{DC}}^{\max} = 1000\,\text{MW} \) in the benchmark scenario, and \( \alpha_t \) represents the normalized load profile obtained from historical records. This approach captures the temporal variability of computing workloads and ensures more realistic representation of data center operation in the system-level analysis.

\subsection{Benchmark system cost result}

We evaluate the impact of data center–based frequency response on system performance through sensitivity analyses that vary two key parameters: the proportion of flexible capacity and the speed of responses.

\subsubsection{Proportion of Flexible Capacity within Data Centers}
The share of total data center load capable of participating in frequency response is varied from 0\% (no participation) to 100\% (full controllability). Greater flexibility generally reduces system operating costs by lowering the need for expensive spinning reserves or thermal ramping during frequency events. However, the marginal benefit declines once flexibility reaches a level sufficient to meet most frequency reserve requirements.

\subsubsection{Response Speed of Data Center FFR}
We also assess the sensitivity of system cost to FFR delivery speed. Faster responses allow frequency deviations to be contained earlier, reducing reliance on slower and costlier primary frequency control from generators. Conversely, delayed responses diminish the effectiveness of data centers as substitutes for fast reserves.

\begin{figure}[h]
\centering
\includegraphics[width=0.5\textwidth]{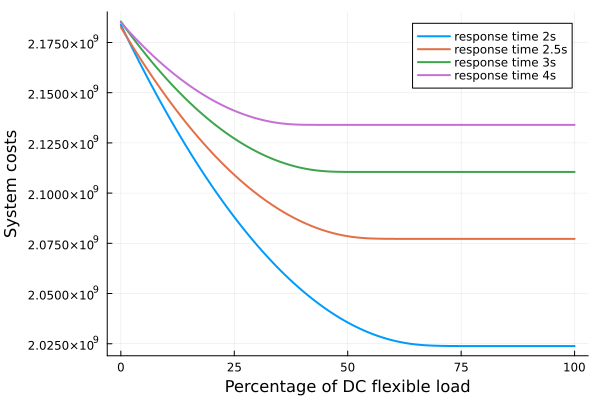}
\caption{Impact of the percentage of DC flexible load on system costs under different FFR response times.}
\label{fig:systemcost} 
\end{figure}

As shown in Fig.~\ref{fig:systemcost}, increasing DC flexibility consistently lowers system costs for all tested response times, with the largest savings achieved at lower flexibility levels (0–50\%). Beyond 75\%, the cost reduction plateaus, reflecting saturation of available demand-side flexibility. Faster response times deliver noticeably greater savings than slower ones, underscoring the importance of response latency in unlocking the full economic value of DC-based FFR.

To quantify the marginal economic benefit of increased data center flexibility, we define the \textit{Marginal Flexibility Value} (MFV) as the system cost reduction per 1\% increase in DC flexible load capacity:

\[
\text{MFV} = \frac{C_{i} - C_{j}}{\phi_{j} - \phi_{i}}
\]

where \( C_i \) and \( C_j \) are the total system costs corresponding to data center flexible load proportions \( \phi_i \) and \( \phi_j \), respectively. The variable \( \phi \in [0, 1] \) denotes the share of the data center load that is responsive to grid frequency signals, expressed as a fraction of the total data center capacity. The unit of MFV is million dollars per 1\% flexible capacity (\$M/\%), representing the cost reduction achieved by enabling one additional percent of data center load to provide frequency response.

\begin{figure}[htbp]
\centerline{\includegraphics[width = 0.5\textwidth]{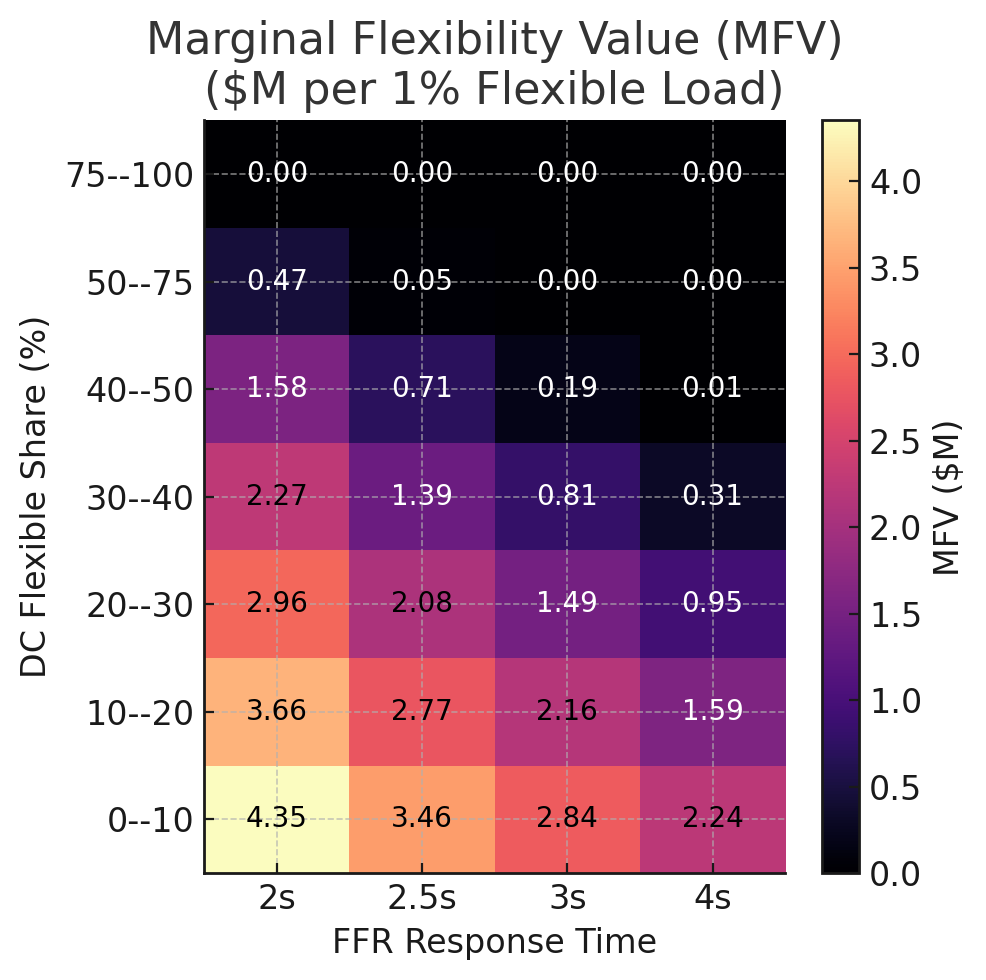}}
\caption{Heatmap of MFV across different DC flexible load shares and FFR response times. Each cell denotes the additional cost saving (\$M) per 1\% increase in flexible load share. Darker colors indicate higher marginal benefit.}
\label{MFV}
\end{figure}

Fig. \ref{MFV} presents the MFV values under different FFR response times. The initial segments (0–25\%) exhibit the highest MFV, particularly at faster response times, indicating that early investments in fast-response flexibility provide the greatest economic return. As flexible load proportion increases, MFV declines, reflecting diminishing marginal savings due to saturation of the system’s frequency reserve needs.

Overall, the results indicate that both the depth (i.e., the proportion of controllable load) and the speed (i.e., response time) of DC-based FFR are key drivers of system cost savings. System operators and policymakers may therefore consider incentivizing not only participation levels but also technological upgrades that shorten response times.

\subsection{2030 Scenario with Increased Data Center Load Penetration}

To evaluate the future implications of growing data center demand, we construct a 2030 scenario based on projected capacity expansion trends. In this scenario, the total data center capacity is increased from 1000\,MW to 2000\,MW, representing a significantly larger share of the system load. To ensure consistency, system-wide demand and renewable generation capacity are also scaled proportionally, maintaining the original wind penetration ratio. This setup enables a fair comparison between present and future system configurations.

Table~\ref{tab:mfv_future_compare} presents the MFV values under current and future data center capacities, assuming a 2\,s response delay. The results highlight two key trends. First, as the flexible share increases, the marginal benefit of additional flexibility gradually declines, reflecting saturation of the system’s frequency reserve requirements. Second, and more importantly, the total economic value of FFR from data centers increases substantially in the 2030 scenario. Early-stage flexibility (0-10\%) yields more than twice the cost reduction per percentage point compared to the current scenario, indicating that as data centers become a larger portion of the system load, their fast-response capability becomes more impactful and valuable from a system-level perspective.

\begin{table}[h]
\centering
\small  
\caption{MFV comparison for 1000\,MW and 2000\,MW DC capacities under 2\,s response delay}
\label{tab:mfv_future_compare}
\renewcommand{\arraystretch}{1.1}

\begin{tabular}{ccc}
\toprule
\multirow{2}{*}{\textbf{DC Flexible Share (\%)}} & \multicolumn{2}{c}{\textbf{DC Capacity}} \\
\cmidrule(lr){2-3}
& 1000 MW & 2000 MW \\
\midrule
0--10     & 4.35 & 9.16 \\
10--20    & 3.66 & 8.38 \\
20--30    & 2.96 & 7.55 \\
30--40    & 2.27 & 6.71 \\
40--50    & 1.58 & 5.86 \\
50--75    & 0.47 & 4.28 \\
75--100   & 0.00 & 0.35 \\
\bottomrule
\end{tabular}

\end{table}

As shown in Table~\ref{tab:windshare}, increasing the proportion of flexible load within data centers not only improves system economics, but also enhances the effective integration of wind energy. In both present (1000\,MW) and future (2000\,MW) data center capacity scenarios, the share of wind power in total generation grows with greater DC-based flexibility. This reflects the improved system capability to accommodate variable renewable generation without violating frequency security constraints.

\begin{table}[h]
\centering
\footnotesize
\caption{Wind power share under different DC flexible load ratios with 2s FFR delay in current and future capacity scenarios}
\label{tab:windshare}
\renewcommand{\arraystretch}{1.1}
\begin{tabular}{ccc}
\toprule
\multirow{2}{*}{\textbf{DC Flexible Share (\%)}} & \multicolumn{2}{c}{\textbf{Wind Power Share (\%)}} \\
\cmidrule(lr){2-3}
& Benchmark scenario & 2030 scenario \\
\midrule
0--10     & 18.3 & 23.5 \\
10--20    & 19.4 & 26.8 \\
20--30    & 20.1 & 28.2 \\
30--40    & 20.3 & 29.3 \\
40--50    & 20.5 & 30.0 \\
50--75    & 20.6 & 30.5\\
75--100   & 20.7 & 30.9 \\
\bottomrule
\end{tabular}
\end{table}

\subsection{Methods Comparison}
To further demonstrate the advantages of the proposed decision tree–based constraint learning (DT-CL) method, we conduct comparative studies against two benchmark linearization approaches: (1) conventional piecewise linear approximation (PLA), and (2) K-means–based regional linearization (KRL). 

The PLA method discretizes nonlinear relationships into manually defined linear segments across input ranges, commonly used for embedding nonlinearity into MILP formulations. The KRL, on the other hand, employs unsupervised clustering to group historical samples into linearly separable regions, followed by local linear fitting within each cluster.

Table~\ref{tab:quant_comparison} provides a quantitative comparison of the three methods applied to the Safe UC problem. All approaches result in an equal number of linear constraints, ensuring a fair comparison in terms of model complexity. The Feasibility Pass Rate represents the proportion of operating points in the validation set (8,760 hourly time steps) where the linearized constraints, when substituted back into the original non-linear (NL) constraint functions, are satisfied. The Average NL Constraint Error measures the mean relative deviation between the linearized and NL constraint values across all validation points, indicating the approximation error introduced by the linearization process. A lower value implies a closer match between the linearized constraints and the original NL representation.

\begin{table}[h]
\centering
\caption{Quantitative Comparison of Linearization Methods in the UC Model}
\label{tab:quant_comparison}
\begin{tabular}{lccc}
\toprule
\textbf{Metric} & \textbf{PLA} & \textbf{K-means} & \textbf{DT-CL (Ours)} \\
\midrule
Linear Constraints & 2,951,904  & 2,951,904  & 2,951,904 \\
MILP Solving Time (s) & 36.72  & 37.23  & \textbf{36.51} \\
System Operation Cost (\$M) & 2298.17  & 2249.32 & \textbf{2183.81} \\
Feasibility Pass Rate (\%) & 97.2 & 98.6 & \textbf{100} \\
Average NL Constraint Error & 0.132 & 0.00561 & \textbf{0.00128} \\
\bottomrule
\end{tabular}
\end{table}

Despite the parity in constraint count, DT-CL achieves the lowest system operation cost, reducing expenditure by 5.0\% relative to PLA and 2.9\% relative to KRL. This improvement stems from its higher linearization fidelity: the Average NL Constraint Error of DT-CL (0.00128) is over two orders of magnitude lower than PLA (0.132) and more than four times lower than KRL (0.00561). The higher accuracy also translates into perfect physical feasibility, as evidenced by a 100\% Feasibility Pass Rate, compared with 97.2\% for PLA and 98.6\% for KRL. In practical terms, this means that DT-CL produces UC schedules that satisfy the original nonlinear nadir frequency constraint in all hourly operating conditions, avoiding the security violations that occur in the benchmark methods.

The MILP solving times are comparable across all methods, with differences within 2\%, indicating that the improved accuracy of DT-CL does not introduce additional computational burden. Taken together, these results demonstrate that DT-CL offers a superior trade-off between accuracy, feasibility, and tractability, making it an effective approach for embedding dynamic security constraints into large-scale UC formulations.

In addition to the linearization-based methods, we also experimented with solving the original nonlinear nadir constraint formulation using a gradient-based optimization approach based on the Adaptive Moment Estimation (ADAM) \cite{kingma2014adam} algorithm—a first-order method widely used in deep learning. However, this approach exhibited prohibitively long convergence times and frequently failed to identify feasible solutions within acceptable tolerances. Due to these practical limitations, it is excluded from the main comparative analysis.

\section{Conclusion}

As data centers continue to expand in scale and energy intensity, their role in modern power systems is becoming increasingly critical. The growing share of data center demand poses challenges for system reliability, but also presents new opportunities for flexible load management. In particular, leveraging data center flexibility for frequency support can enhance system stability while reducing operational costs. However, effectively integrating this flexibility into power system scheduling requires tractable models that respect both physical limits and data center operational constraints.

To address this challenge, we propose a Safe UC framework that embeds data-driven frequency security constraints combining data center frequency response into a traditional UC model. The core innovation lies in a constraint learning module based on decision tree classification, trained on historical frequency response data. By translating the learned structure into piecewise linear constraints, the method produces interpretable, MILP-compatible safety constraints that capture real-world system behavior under uncertainty.

We evaluate the system-level impact of flexible data center participation through two representative scenarios: a present-day benchmark and a projected 2030 future with doubled data center capacity. Results show that increasing the share of fast-responding flexible load within data centers leads to significant reductions in total system cost. However, this benefit is subject to diminishing returns, as revealed by the proposed Marginal Flexibility Value metric. For example, under a 2-second response delay, the MFV reaches \$4.35M per additional 1\% of flexible load in our benchmark scenario, rising to \$9.16M in the 2030 scenario.

Beyond economic savings, enhanced data center flexibility also facilitates greater integration of renewable energy. In the 2030 case, the share of wind generation increases from 23.5\% to 30.9\% as DC flexibility expands from 0\% to 100\%, highlighting a strong co-benefit between frequency-secured scheduling and decarbonization goals.

In summary, the Safe UC framework offers a scalable and interpretable approach to co-optimizing economic dispatch, frequency security, and renewable integration. Future work will extend this framework to incorporate stochastic uncertainty, geographically distributed data center clusters, and intelligent workload scheduling for dynamic load shaping.


\bibliographystyle{IEEEtran}
\bibliography{bib/IEEEexample}

\end{document}